\documentclass[reprint,superscriptaddress,amsmath,amssymb,aps,pre]{revtex4-2}
\usepackage{graphicx}
\usepackage{dcolumn}
\usepackage{upgreek}
\usepackage{bm}
\usepackage{amsmath}
\usepackage[font=small]{caption}
\usepackage[labelformat=empty, position=top]{subcaption}
\usepackage[export]{adjustbox}
\usepackage{blindtext}
\usepackage{amssymb}
\usepackage{newunicodechar}
\usepackage{nameref}
\usepackage{xcolor}
\begin{document}

\title{Multi-Phase Locking Value: A Generalized Method for Determining Instantaneous Multi-frequency Phase Coupling}

\author{Bhavya Vasudeva}
  \affiliation{Indian Statistical Institute, Kolkata, West Bengal 700108, India}

\author{Runfeng Tian}
  \affiliation{Stephenson School of Biomedical Engineering, The University of Oklahoma, Tulsa, Oklahoma-74135, USA}

\author{Dee H. Wu}
  \affiliation{Department of Radiological Sciences, The University of Oklahoma Health Sciences Center, Oklahoma City, Oklahoma 73104, USA}  

\author{Shirley A. James}
  \affiliation{Department of Rehabilitation Sciences, College of Allied Health, The University of Oklahoma Health Sciences Center, Oklahoma City, Oklahoma 73117, USA} 

\author{Hazem H. Refai}
  \affiliation{Department of Electrical and Computer Engineering, The University of Oklahoma, Tulsa, Oklahoma-74135, USA}
  
\author{Fei He}
  \affiliation{Centre for Computational Science and Mathematical Modelling,, Coventry University, Coventry CV1 2JH, UK}
  
\author{Yuan Yang}
  \email{yuan.yang-2@ou.edu}
  \affiliation{Stephenson School of Biomedical Engineering, The University of Oklahoma, Tulsa, Oklahoma-74135, USA}
  \affiliation{Department of Physical Therapy and Human Movement Sciences, Feinberg School of Medicine, Northwestern University, Chicago, Illinois-60611, USA}

\begin{abstract}
\noindent
Background: Many physical, biological and neural systems behave as coupled oscillators, with characteristic phase coupling across different frequencies. Methods such as $n:m$ phase locking value (where two coupling frequencies are linked as: $mf_1=nf_2$) and bi-phase locking value have previously been proposed to quantify phase coupling between two resonant frequencies (e.g. $f$, $2f/3$) and across three frequencies (e.g. $f_1$, $f_2$, $f_1+f_2$), respectively. However, the existing phase coupling metrics have their limitations and limited applications. They cannot be used to detect or quantify phase coupling across multiple frequencies (e.g. $f_1$, $f_2$, $f_3$, $f_4$, $f_1+f_2+f_3-f_4$), or coupling that involves non-integer multiples of the frequencies (e.g. $f_1$, $f_2$, $2f_1/3+f_2/3$).\\ 
New methods: To address the gap, this paper proposes a generalized approach, named multi-phase locking value (M-PLV), for the quantification of various types of instantaneous multi-frequency phase coupling. Different from most instantaneous phase coupling metrics that measure the simultaneous phase coupling, the proposed M-PLV method also allows the detection of delayed phase coupling and the associated time lag between coupled oscillators.\\
Results: The M-PLV has been tested on cases where synthetic coupled signals are generated using white Gaussian signals, and a system comprised of multiple coupled R\"ossler oscillators, as well as a human subject dataset. Results indicate that the M-PLV can provide a reliable estimation of the time window and frequency combination where the phase coupling is significant, as well as a precise determination of time lag in the case of delayed coupling. This method has the potential to become a powerful new tool for exploring phase coupling in complex nonlinear dynamic systems.  
\end{abstract}


\maketitle
\section{Introduction}

Complex systems such as the human brain behave as a series of oscillators with their instantaneous phases dynamically coupled over multiple frequency bands \cite{varela2001brainweb,breakspear2017dynamic,canolty2010functional,jensen2007cross,he2021}. Sheremet \textit{et al}. \cite{Sheremet} use quadratic nonlinearity to detect cross-frequency coupling between theta and gamma \textcolor{black}{waves} in the hippocampus. Recent works focus on reconstructing coupling functions \cite{Stankovski} and estimating the phase oscillator model \cite{Onojima} using real data. The phase and amplitude dynamics of large nonlinear systems of heterogeneous, globally coupled oscillators \cite{waishinglee} and non-identical damped harmonic oscillators \cite{Cudmore} have also been studied.

Methods such as $n:m$ phase locking value (PLV) \cite{ermentrout1981n}, bi-phase locking value (bPLV) \cite{bplv} and their variants \cite{schelter2006partial,vinck2011improved,vahabi2015online} have previously been proposed to detect and quantify different types of phase coupling. The $n:m$ PLV measures phase coupling between two resonant frequencies when $n$ cycles of one oscillatory signal are phase locked to $m$ cycles of another oscillatory signal, i.e., $m\phi(f_n,t) - n\phi(f_m,t) \leq \epsilon$ \cite{ermentrout1981n, mossbook} ($\phi(f,t)$ is the instantaneous phase at frequency $f$ and time point $t$, two resonant frequencies $f_n$ and $f_m$ are linked as $f_n:f_m = n:m$, and $\epsilon$ denotes a small constant). The bPLV quantifies quadratic phase coupling among three frequencies, where a pair of frequencies, $f_1$ and $f_2$ are coupled to a third frequency $f_3 = f_1+f_2$ or $f_1-f_2$, i.e.,  $\phi(f_1,t)\pm\phi(f_2,t)-\phi(f_3,t) \leq \epsilon$ \cite{bplv}. 

However, phase coupling could be shown in more complicated patterns involving more than three frequencies  (e.g. $f_1$, $f_2$, $f_3$, $f_4$, $f_1+f_2+f_3-f_4$) as well as their non-integer multiples (e.g. $f_1$, $f_2$, $2f_1/3+f_2/3$), which cannot be detected or quantified by using the conventional phase coupling metrics, such as $n:m$ PLV~\cite{ermentrout1981n} and bPLV~\cite{bplv}. A novel measure called multi-spectral phase coherence (MSPC) has been recently developed by Yang and colleagues to provide a generalized approach for quantifying integer multi-frequency phase coupling \cite{mspc}. This method has been applied to the human nervous system to advance our understanding of nonlinear neuronal processes and their functions in movement control \cite{yang2016nonlinear,yang2018unveiling} and sensory perception \cite{gordon2019expectation}. The MSPC is a straightforward extension of bPLV based on high-order spectra \cite{nikias1993signal}; however, it does not cover either the non-integer multi-frequency phase coupling (e.g. $f_1$, $f_2$, $2f_1/3+f_2/3$) or non-integer resonant coupling (e.g. 2:3 coupling \cite{langdon2011multi} revealed by $n:m$ PLV) problems.

Thus, this paper aims to introduce a more generalized approach, namely multi-phase locking value (M-PLV), that integrates the concepts of MSPC and $n:m$ PLV to allow the detection and quantification of various types of phase coupling, including integer and non-integer, multi-frequency and resonant phase coupling. The proposed M-PLV provides us with a tool to explore the unreported non-integer multi-frequency phase coupling that has never been captured by existing phase coupling methods. Furthermore, different from commonly used instantaneous phase coupling metrics, the proposed method also allows the detection of delayed phase coupling and the associated time lag between coupled oscillators. We tested M-PLV on two scenarios where synthetic coupled signals are generated using white Gaussian signals, and a system comprised of multiple coupled R\"ossler oscillators. \textcolor{black}{The real application of M-PLV was demonstrated in a EEG-EMG data dataset recorded from human subjects during a motor task \cite{tian2021assessing}}. 

The rest of this paper is organized as follows: Section 2 describes M-PLV, Section 3 summarizes the experiments used to validate the method, Section 4 presents the results and discussion, and Section 5 concludes the paper.

\section{Multi-phase Locking Value (M-PLV): Theory and Calculation}
The proposed M-PLV is generalized approach that integrates the concept of MSPC \cite{mspc} and $n:m$ PLV \cite{ermentrout1981n}. It not only provides us with a formulated mathematical description for the phase coupling problems separately described by MSPC and $n:m$ PLV, but also permits the detection and quantification of non-integer multi-frequency phase coupling that cannot be assessed by using existing phase coupling methods. 

\subsection{M-PLV}
The MSPC considers the case where multiple input frequencies $f_1,f_2, ..., f_L$ are coupled to an output frequency $f_{\Sigma}$ based on an integer combination, such that $f_{\Sigma} = \sum_{l=1}^{L}m_lf_l$, $m_l \in \mathbb{N}$:
\begin{equation}
\left( \sum_{l=1}^{L} m_l\phi(f_l,t)\right) -\phi(f_{\Sigma},\,t) \leq \epsilon
\end{equation}

\textcolor{black}{The formula for MSPC is given by:
\begin{align}
&MSPC(f_1,\, f_2,\, ...,\, f_L;\, m_1,\, m_2,\, ...,\, m_L;\,t) =& \nonumber\\ 
&\left|\frac{1}{K} \sum_{k=1}^{K} \exp \left(j\left( \left( \sum_{l=1}^{L} m_l\phi_{k}(f_l,\,t)\right) -\phi_{k}(f_{\Sigma},\,t)\right)\right)\right|&
\end{align}
}

The MSPC does not cover the case where non-integer multiples of input frequencies are coupled to the output frequency. To address this gap, the proposed M-PLV generalizes the relation between frequencies as $nf_{\Sigma} = \sum_{l=1}^{L}m_lf_l$ or $f_{\Sigma}=\sum_{l=1}^{L}\frac{m_l}{n}f_l$ \textcolor{black}{($L$ is a finite integer)}. It can be seen that although $m_l,\,n$ are integers, their ratio can give \textcolor{black}{rational numbers}. This idea is in line with the concept of $n:m$ PLV \cite{ermentrout1981n}, but allows assessment of phase coupling between multiple input frequencies and one targeted output frequency. 

\textcolor{black}{Moreover, there may exist a delay $\tau$ in the system between the input and the output, such that the coupling can be detected only after this delay has been compensated by aligning the indices of all the instantaneous phases. Incorporating these factors, the proposed M-PLV aims to detect and quantify a more generalized phase coupling phenomenon that can be described as:}
\begin{equation}
\left( \sum_{l=1}^{L} m_l\phi(f_l,t-\tau)\right) -n\phi(f_{\Sigma},\,t) \leq \epsilon
\end{equation}

Based on this theoretical definition \textcolor{black}{and the formulae used by other methods to quantify phase coupling}, the formula of M-PLV ($\Psi$) is given as follows for the calculation:
\begin{align}
\label{MPLV}
&\Psi(f_1,\, f_2,\, ...,\, f_L;\, m_1,\, m_2,\, ...,\, m_L,\,n;\,t,\,\tau) =& \nonumber\\ 
&\left|\frac{1}{K} \sum_{k=1}^{K} \exp \left(j\left( \left( \sum_{l=1}^{L} m_l\phi_{k}(f_l,\,t-\tau)\right) -n\phi_{k}(f_{\Sigma},\,t)\right)\right)\right|&
\end{align}

where $K$ is \textcolor{black}{a finite integer} number of observations, $\phi_{k}(f_l,t)$ is an instantaneous input phase at the $k^{th}$ observation, which can be obtained from the Hilbert transform of narrowband filtered time series with the spectrum centered at frequency $f_l$ \cite{hilbertbb}.

\subsection{Detecting significant M-PLV}
In order to detect the time window and frequency at which phase coupling is significant, a reference threshold value of M-PLV is required. For this purpose, the 95\% significance threshold is obtained by a Monte Carlo simulation~\cite{mspc}, which is a generally acceptable confidence level for determining statistical significance~\cite{wang2004exact}. The Monte Carlo is a typical method to show the significance of cross-spectral based analysis such as coherence and phase coupling~\cite{pardo2012spectral}. The null hypothesis is that the phase difference $\Delta \phi(t;k)$ is completely random so that the cyclic phase difference $\Delta \upphi(t;k)=\Delta \phi(t;k)$ mod $2\pi$ will be uniformly and randomly distributed in the interval $[0,2\pi]$. The cyclic phase difference is used here because the phase returned by taking the
inverse of a sinusoid is cyclic/periodic with period $2\pi$. The M-PLV corresponding to other frequency combinations for all instants $t$ as well as those corresponding to the combination of interest for the instants $t'=t-t_c$ ($t_c$ is the estimated coupling window) are taken as surrogate data of uniformly and randomly distributed phase values of $\Delta \phi(t;k)$. This procedure is repeated $N$ times (typically $N = 1000$ is sufficient for a reliable Monte Carlo simulation for phase coupling measures~\cite{mspc}) to obtain the statistical distribution of M-PLV values for a given number of observations, which is determined by the experimental design or available real data. Then, the threshold is determined as the minimum value greater than 95\% of the sum of all the values in the distribution.

\subsection{Delay Estimation}
In order to estimate the delay $\tau$, the M-PLV for different values $\tau_i$ within a given range \textcolor{black}{is} calculated. The value of $\tau_i$ corresponding to the maximum value of M-PLV is the estimated delay $\hat{\tau}$ of the system.

\section{Experiments}

We tested M-PLV on two scenarios where synthetic coupled signals are generated (1) using white Gaussian signals alone, as well as (2) from a system comprised of multiple coupled R\"ossler oscillators. In these simulations, the sampling frequency is 1 kHz. Noteworthy, the numerical values used in the simulations are just example values for testing the proposed method. In real applications, different numerical values could be used based on real experimental data. For example, we applied M-PLV to check 1:1 (integer) and 2:1 (1/2 non-integer) coupling and estimated delay between electroencephalography (EEG) and electromyography (EMG) signals during a motor task to demonstrate a real application of M-PLV (see Section~\ref{realdata}), where the numerical values are from the real data obtained in a human subject experiment~\cite{tian2021assessing}.

\subsection{Coupled white Gaussian signals}
In this case, $x(t)$ and $y(t)$ are two independent white Gaussian signals (zero mean and unit variance). The synthetic signal, $y_{c}(t)$ is generated as follows:
\begin{align}
\label{ycoupled}
     y_{c}(t) = y(t) &- y(f_{\Sigma}, t_c) & \nonumber\\
     &+ \frac{x^{|m_1|}(f_1, t_c)x^{|m_2|}(f_2, t_c)}{A_x^{|m_1|}(f_1, t_c)A_x^{|m_2|}(f_2, t_c)}A_y(f_{\Sigma}, t_c)
 \end{align}

where $t$ is in the range of [0.001,10] s, $x(t_c)$ represents $x(t)$ in the phase coupling time window $t_c = [2.501, 7.5]$ s. $x(f_1, t_c)$ is a narrowband signal with the spectrum centered at frequency $f_1$, which is obtained after $x(t_c)$ is passed through a Butterworth band-pass filter \cite{Chua1987LinearAN} centered at frequency $f_1$ (bandwidth: 2 Hz, $6^{th}$ order). $A_x(f_1, t_c)$ is the envelope of the Hilbert transform of $x(f_1, t_c)$. In order to eliminate the effect of filter on the signal phase, zero-phase shift filter (Matlab function: filtfilt.m) is used in this study. The normalization of the signal $x(f_1, t_c)$ by its envelope $A_x(f_1, t_c)$ prevents abrupt changes in its amplitude. 

In these designed signals, there is phase coupling between $y_{c}(t)$ and $x(t)$ in the time interval $t_c$, following the rule $f_{\Sigma}=m_1f_1+m_2f_2$, serving as the ground truth in this ``white'' box problem for testing the M-PLV for integer ($n=  1$) multi-frequency phase coupling with zero delay ($\tau= 0$).  

In order to check for the phase coupling between $x(t)$ and $y_c(t)$, M-PLV is calculated based on Eq.~(\ref{MPLV}), and the set of input frequencies includes $f_1$ and $f_2$.

\subsection{Coupled R\"ossler oscillators}
In this case, $y(t)$ is white Gaussian signal, while $x_i(t)$ are obtained from a system comprised of coupled R\"ossler oscillators \textcolor{black}{in the chaotic regime}, which consists of $N-1$ independent oscillators coupled to the $N^{th}$ oscillator. The system is characterized by the following equations:
\begin{align}
   \dot{x}_i &= 
	-\left(\sum_{j=1}^{N-1}\frac{m_j \omega_j}{n}\right)  x_i - z_i + \varepsilon_i  \left(\sum_{j=1}^{N-1} \frac{m_jx_j}{n} - x_i\right) \\ 
\label{eq:rossler1}
\dot{u}_i &=\omega_i  x_i + a u_i\\
\dot{z}_i &= c + z_i  (x_i - b) 
\label{eq:rossler2}
\end{align}
    
where $\varepsilon_i=0$ for $i<N$ \textcolor{black}{(N is a finite integer)} and $\omega_j=2\pi f_j$. These coupled oscillators are designed to mimic a multiple-input-single-output (MISO) system. In this case, Eq.~(\ref{ycoupled}) can be generalized to include a larger number of signals coupled at different frequencies, so that $nf_{\Sigma}=\sum_{l=1}^{N}m_lf_l$ and the coupled signal can be obtained as follows: 
\begin{align}
      y_{c}(t) &= y(t) - y(nf_{\Sigma}, t_c) & \nonumber \\
      &+ A_y(nf_{\Sigma}, t_c)\left(\frac{x_1^{|m_1|}(f_1, t_c)}{A_{x_1}^{|m_1|}(f_1, t_c)}...\frac{x_N^{|m_N|}(f_N, t_c)}{A_{x_N}^{|m_N|}(f_N, t_c)}\right)
\end{align}

where $x_i(f_j, t_c)$ is obtained after $x_i(t_c)$ is passed through a Butterworth band-pass filter \cite{Chua1987LinearAN} centered at frequency $f_j$ (bandwidth: 2 Hz, $6^{th}$ order). In order to introduce a delay $\tau$ in the system, $t_c$ can be replaced by $t_c-\tau$ in the above equation. The coupling is evaluated between $x_N(t)$ and $y_c(t)$ by calculating the M-PLV according to Eq.(\ref{MPLV}).


The 95\% significance threshold and delay $\tau$ can be estimated through the procedure described in Section 2.2 and 2.3.

\subsection{EEG-EMG dataset} \label{realdata}
The real application of the proposed method is demonstrated in the EEG and EMG data from four healthy participants that were recorded in a previous study at Northwestern University, Chicago, USA \cite{tian2021assessing}. In this previous study, the participants were recruited with written informed consent and permission of the Northwestern University institutional review board. 
Participants were seated with tested arm positioned with 85$^{\circ}$ shoulder abduction, 45$^{\circ}$ shoulder flexion and 90$^{\circ}$ elbow flexion in a Biodex pedestal. Maximum voluntary torque (MVT) of the shoulder abduction (SABD) was measured at the beginning of the experiment for each participant. After that, the participants were asked to lift the tested arm and hold for 10 seconds with $40\%$ of SABD MVT for each trial. In total, the trials were repeated for 25 times.
32-channel EEG (Biosemi, Inc, Active II, Amsterdam, the Netherlands) was recorded using 10/20 recording system. The EMG from muscle activity at Intermediate Deltoid of the tested arm was recorded simultaneously during the experiment. The brain and muscles are coupled during the
movement task since the brain controls/communicates with the muscles \cite{tian2021assessing,yang2018unveiling}. Thus, this dataset is suitable to test the proposed method. The C3 (if the tested arm is right arm) or C4 (if the tested arm is left one) channel of EEG was used in this project to compute the coupling between EEG and EMG. These EEG channels are used since they are over brain regions in the primary motor cortex controlling arm movements~\cite{yang2016nonlinear2,weersink2019eeg}. Both EEG and EMG were sampled at 2048 Hz.

\section{Results and Discussion}
\subsection{Coupled white Gaussian signals: integer multi-frequency phase coupling with zero delay}
\begin{figure*}[tb]
    \centering
    \includegraphics[width=0.9\linewidth]{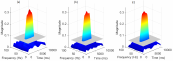}
    \caption{M-PLV as a function of time and frequency for K = (a) 500, (b) 750 and (c) 900.}
    \label{fig1}
\end{figure*}

\begin{figure}[tb]
    \centering
    \includegraphics[width=1.0\linewidth]{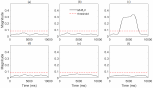}
    \caption{M-PLV for K = 600 as a function of time (unit: ms) for the set of frequencies (a) 3, (b) 39, (c) 45, (d) 55, (e) 71, and (f) 87 Hz.}
\label{fig2}
\end{figure}

The results are shown for $f_1=29$ Hz, $f_2=13$ Hz, $m_1=2$, and $m_2=-1$, so that $f_{\Sigma}=2 \times 29 - 1\times 13=45$ Hz. Fig.~\ref{fig1} shows M-PLV plotted as a function of time and frequency for varying numbers of epoches $K\, (K = 500,\, 750,\, 900)$. M-PLV is calculated for all possible \textcolor{black}{linear} combinations of the frequencies $f_1=29$ and $f_2=13$ Hz \textcolor{black}{with integral weights} to examine whether the significant M-PLV is only detected on the target frequency 45 Hz rather than other frequencies. It is observed that M-PLV shows significant values for $f_{\Sigma} = 45 Hz$ in the time window $t'_c\sim t_c= [2.501, 7.5]$, i.e.,  the interval $t'_c=[2.492, 7.511]$ s, $[2.421, 7.461]$ s, and $[2.431, 7.6]$ s for $K = 500,\, 750,$ and $900$, respectively. The error of time window estimation can be defined as the difference between $t'_c$ and $t_c$ and divided by the window size. The errors are below 5~\% for all tested $K$ values. To further demonstrate the performance of M-PLV, Fig.~\ref{fig2} shows a few of example plots of M-PLV for $K = 600$ for some possible combination frequencies of $f_1=29$ and $f_2=13$ Hz  (e.g.  $29 - 2 \times 13 = 3$ Hz, $0\times29 + 3\times 13 = 39$ Hz, etc). Significant M-PLV is only detected at the targeted frequency $f_{\Sigma} = 45 Hz$ within the coupled time window. 

\subsection{Coupled R\"ossler oscillators: integer and non-integer multi-frequency phase coupling with a delay}\label{simu-delay}
In these simulations, we set $K = 400$, $N = 3$, and the parameters of the coupled R\"ossler oscillators (Eq.~(\ref{eq:rossler1})~-~(\ref{eq:rossler2})) as $a = 0.15$, $c = 0.2$, $b = 10$, and $\varepsilon_N=0.1$. Noteworthy, the proposed method is able to work with larger $N$. However, without loss of generality, here $N = 3$ has already shown the capability of the proposed method as detailed below.

\begin{figure}[tb]
\centering
\includegraphics[width=1.0\linewidth]{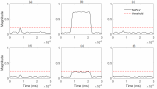}
\caption{M-PLV for the first set of values of the coupled R\"ossler oscillators, as a function of time, for the set of frequencies (a) 1, (b) 7, (c) 9, (d) 11, (e) 13, and (f) 15 Hz. }
\label{fig3}
\end{figure}

\begin{figure}[tb]
\centering
\includegraphics[width=1.0\linewidth]{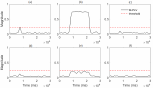}
\caption{M-PLV for the second set of values of the coupled R\"ossler oscillators, as a function of time, for the set of frequencies (a) 1, (b) 7, (c) 9, (d) 11, (e) 13, and (f) 15 Hz. }
\label{fig4}
\end{figure}

To demonstrate the performance of the method for integer ($n=1$) multi-frequency phase coupling with zero delay ($\tau=0$) in a MISO system, the oscillators are simulated for 80 seconds and two sets of 30\,000 samples are obtained from the simulated signals, with $t=[10.001, 40]$ s, $t_c=[17.501, 32.5]$ s for the first set and $t=t+40$ s, $t_c=t_c+40$ s $= [57.501, 72.5]$ s for the second set. In this case, $f_1=3$ Hz, $f_2=5$ Hz, $m_1=-1$, $m_2=2$, so that $f_{\Sigma}= -1\times3 + 2 \times 5=7$ Hz. M-PLV is calculated for possible combinations of the frequencies $f_1=3$ Hz, $f_2=5$ Hz to examine whether the significant M-PLV is only detected on the target frequency 7 Hz rather than others (e.g. $2 \times 3 - 5 = 1$, $2 \times 3 + 5 = 11$, etc.). Fig.~\ref{fig3} and Fig.~\ref{fig4} show M-PLV for the first and second time set, respectively. The coupling is detected in the time window $t'_c=[17.383, 32.286]$ s (error: 2.2\%. Let the coupling interval be $t_c=[t1,t2]$ and the estimated coupling interval be $t_c’=[t1’,t2’]$. Then, the estimated error is given by $100*(|t1-t1’|+|t2-t2’|)/(t2-t1)$ for the first set and $t'_c=[57.484, 72.279]$ s \textcolor{black}{(error: 1.6\%)} for the second set. 

\begin{figure}[tb]
\centering
\includegraphics[width=0.9\linewidth]{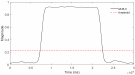}
\caption{M-PLV as a function of time for $n=5$ (non-integer multiples) at frequency 4 Hz.}
\label{fig5}
\end{figure}

To demonstrate the performance of the method for non-integer multi-frequency phase coupling, the procedure is repeated for another case where $f_1=7$ Hz, $f_2=13$ Hz, $m_1=1$, $m_2=1$, and $n=5$ so that $f_{\Sigma}=\frac{(1\times7+1\times13)}{5}=4$ Hz. Also, $t=[10.001, 40]$ s and $t_c=[17.501, 32.5]$ s. Fig.~\ref{fig5} shows the results obtained for $f_{\Sigma}=4$ Hz. Using the 95\% significance threshold, $t'_c=[17.295, 32.444]$ s \textcolor{black}{(error: 1.7\%)}.

\begin{figure}[tb]
\centering
\includegraphics[width=0.9\linewidth]{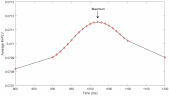}
\caption{Average M-PLV as a function of delay $\tau$. The local maxima occurs at  $\hat{\tau}=1.02$ s in this case.}
\label{fig6}
\end{figure}

To demonstrate the performance of the method for delay estimation, the synthetic signal is generated after $\tau$ is set as $1$ s. In this case, $t=[10.001, 40]$ s, $t_c=[17.501, 32.5]$ s, $f_1=29$ Hz, $f_2=13$ Hz, $m_1=2$, and $m_2=-1$, so that $f_{\Sigma}=45$. Fig.~\ref{fig6} shows the average M-PLV obtained for varying $\tau_i$. The estimated local maxima over 10 such simulations is $\hat{\tau}=0.994\pm0.0568$ s, with an average error less than $5$\%.

\subsection{Phase coupling between brain and muscle activities with a delay: a real application}\label{real-delay}
\textcolor{black}{For this case, the signals were first low-pass filtered using a sixth order Butterworth filter with cutoff frequency 256 Hz and downsampled to frequency 512 Hz. Then, we calculated 1:1 M-PLV for frequencies in the range 14-40 Hz, averaged over 25 trials for 4 subjects, for different values of the delay $\tau$. On comparing average M-PLV for various values of $\tau$, we get $\hat \tau=$ 25.4 ms. Fig. 7 shows M-PLV as a function of time for various frequencies. The estimated time delay is \textcolor{black}{in} line with the nerve conduction delay from the brain to the muscles reported in the previous experimental studies \cite{perenboom2015evidence,witham2011contributions}.}

\begin{figure}[tb]
\centering
\includegraphics[width=1.0\linewidth]{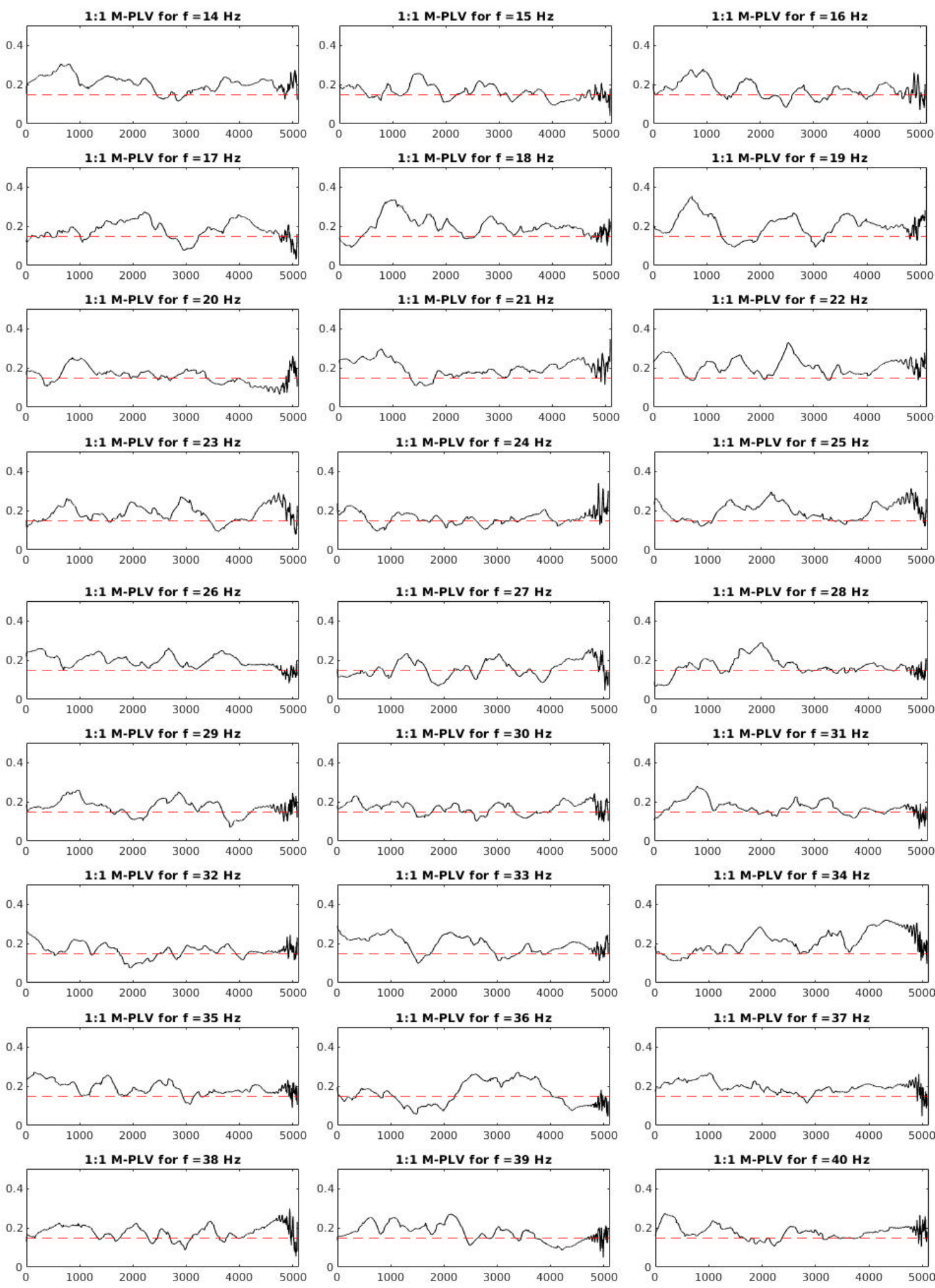}
\caption{\textcolor{black}{1:1 M-PLV for the EEG-EMG signals, as a function of time for frequencies between 14-40 Hz.}}
\label{fig7}
\end{figure}

\textcolor{black}{Next, we checked for 2:1 coupling between EEG and EMG signals with the same time delay. This is because, in healthy participants, nonlinear coupling is generated in the same motor descending pathway as the linear coupling \cite{yang2018unveiling}.  Fig. 8 shows results where EEG at 20 and 26 Hz is coupled to EMG at 10 and 13 Hz.}

\begin{figure*}[tb]
\centering
\includegraphics[width=0.8\linewidth]{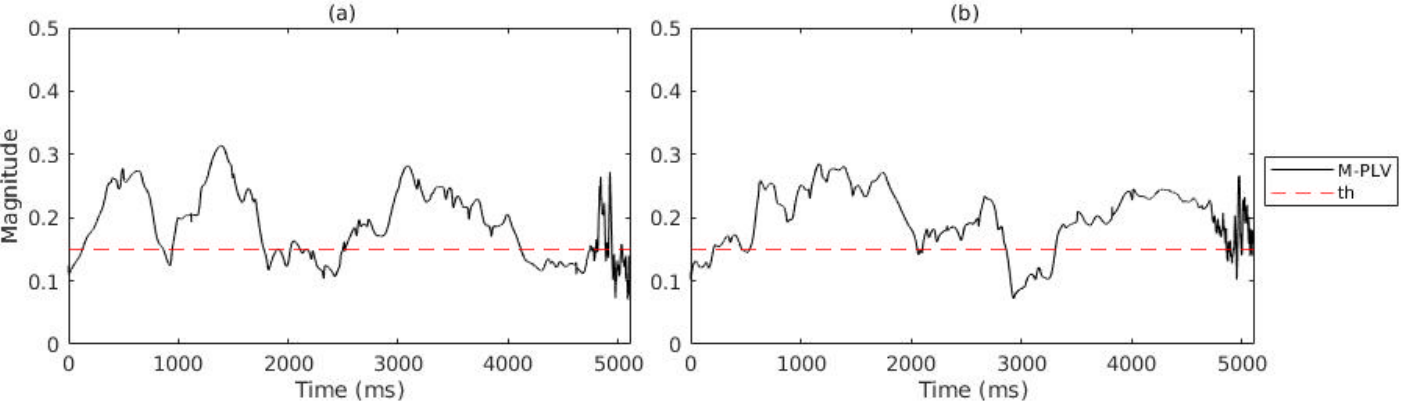}
\caption{\textcolor{black}{2:1 M-PLV for the EEG-EMG signals, as a function of time for frequencies (a) 20 and (b) 26 Hz.}}
\label{fig8}
\end{figure*}

\textcolor{black}{Although a continuous shoulder abduction torque was generated during the motor control task, both linear and nonlinear parts do not show continuous coupling. This is likely related to the discontinuous firing patterns of neurons in the motor descending pathway which may be associated with the excitatory and inhibitory processes of the continuous motor command \cite{staude2000discontinuous}.}   

\subsection{Comparison of M-PLV, MSPC, bPLV, and n:m PLV}
When time delay $\tau = 0$, the proposed method can be used for detecting and quantifying the simultaneous multi-frequency phase coupling. Additionally, if $n = 1$, M-PLV is further degraded to MSPC, for measuring simultaneous integer multi-frequency phase coupling: 

\begin{align}
\label{MSPC}
&MSPC(f_1,\, f_2,\, ...,\, f_L;\, m_1,\, m_2,\, ...,\, m_L;\,t) = & \nonumber\\ 
&\left|\frac{1}{K} \sum_{k=1}^{K} \exp \left(j\left( \left( \sum_{l=1}^{L} m_l\phi_{k}(f_l,\,t)\right) -\phi_{k}(f_{\Sigma},\,t)\right)\right)\right|&
\end{align}

Noteworthy, bPLV \cite{bplv} is basically a special form of MSPC or M-PLV when the interest is in determining quadratic phase coupling: 

\begin{align}
\label{bPLV}
& bPLV(f_1,\, f_2;\,t) & \nonumber\\
& = \left|\frac{1}{K} \sum_{k=1}^{K} \exp \left(j\left(\left( \phi_{k}(f_1,\,t) + \phi_{k}(f_2,\,t)\right) -\phi_{k}(f_{1}+f_{2},\,t)\right)\right)\right|
\end{align}

When $L = 1$, M-PLV can also be degraded to $n:m$ PLV\cite{ermentrout1981n}:

\begin{align}
& PLV_{n:m}(f_n,\, f_m;\,m,\,n;\,t) & \nonumber\\
& =\left|\frac{1}{K} \sum_{k=1}^{K} \exp \left(j\left(m\phi_{k}(f_n,\,t) -n\phi_{k}(f_m,\,t)\right)\right)\right|
\end{align}

As such, M-PLV not only allows the detection and quantification of delayed coupling, non-integer and integer multi-frequency coupling, but also provides a generic mathematical framework that can accommodate all common forms of phase coupling in the existing literature. Noteworthy, simultaneous phase coupling measures MSPC, bPLV, and n:m PLV are not able to correctly detect the delayed coupling (showing non-significant values) such as the cases shown in Section~\ref{simu-delay} (simulation) and~\ref{real-delay} (real data) since there is no time delay $\tau$ in their definitions. The comparison of M-PLV, MSPC, bPVL and n:m PLV is summarized in Table~\ref{table1} \\

\begin{table*}[tb]
\caption{Comparison of different phase coupling methods\label{table1}}
 \begin{tabular}{c c c} 
 \hline 
 Methods & Type of phase coupling & Type of dynamic coupling\\ 
 \hline 
 M-PLV & All multi-frequency coupling & Coupling with/without delays \\  
 MPSC~\cite{mspc} & Integer multi-frequency coupling only & Coupling without delays only \\
 bPLV~\cite{plv}  & Quadratic coupling only & Coupling without delays only \\
 m:n PLV~\cite{ermentrout1981n} & Two frequency coupling only &  Coupling without delays only \\
 \hline
 \end{tabular}
\end{table*}

\section{Conclusion}
In this paper, a new method for quantifying multi-frequency phase coupling has been proposed. This method addresses the limitation of existing approaches that only allow the detection of coupling between two resonant frequencies (i.e. $n:m$ PLV) or quadratic coupling between three frequencies (i.e. bPLV). The M-PLV allows us to quantify various types of phase coupling, including both integer and non-integer phase coupling across multiple frequencies, so as to permit the exploration of more complicated, even unreported phase coupling phenomena in the real world. Simulation studies have been performed on synthetic coupled signals generated using white Gaussian signals, and a complex system comprised of multiple coupled R\"ossler oscillators. \textcolor{black}{We also tested our approach for a real-time application to check neural coupling between electrical brain (EEG) and muscle (EMG) signals.} Our results suggest that the proposed method can achieve a reliable estimate of the frequency combination as well as the time window during which phase coupling is present. Furthermore, this method can be used for a precise estimation of the delay between the input and the output when delayed phase coupling is present between the oscillators. \textcolor{black}{This method has the potential to become a powerful new tool for exploring phase coupling in complex nonlinear dynamic systems such as the human motor system}.

\section*{Declarations}
\subsection*{Funding}
This work was supported by NIH R21HD099710 and P20GM121312, OCAST HR21-164-1 and NSF RII Track-2 FEC 1539068. B. Vasudeva received stipend from S. N. Bose Scholars Program 2019.
\subsection*{Conflicts of interest/Competing interests}
Authors claim that they do not have any conflicts of interest.

\newpage
\bibliography{mybibfile_1}

\end{document}